\documentclass[12pt]{article}
\usepackage{graphicx}

\def\de{\delta}

\newcommand{\De}{\Delta}
\newcommand{\SSS}{\cal S}
\newcommand{\bn}{{\bf n}}

\newcommand{\bk}{{\bf k}}
\newcommand{\be}{\begin{equation}}
\newcommand{\ee}{\end{equation}}

\newsavebox{\sboxpubnumber}
\newsavebox{\sboxpubdate}
\newcommand{\pubdate}[1]{\begin{lrbox}{\sboxpubdate}{#1}\end{lrbox}}

\newcommand{\Title}[1]{\begin{center} {\Large #1 } \end{center}}
\newcommand{\Author}[1]{\begin{center}{ \sc #1} \end{center}}
\newcommand{\Address}[1]{\begin{center}{ \it #1} \end{center}}

\newenvironment{Abstract}{\begin{quotation}  }{\end{quotation}}
\newenvironment{Presented}{\begin{quotation} \begin{center}
             PRESENTED AT\end{center}\bigskip
      \begin{center}\begin{large}}{\end{large}\end{center}
      \end{quotation}}
\newcommand{\Acknowledgements}{\bigskip  \bigskip \begin{center} \begin{large}
             \bf ACKNOWLEDGEMENTS \end{large}\end{center}}
\begin{document}

%%%%%%%%%%%%%%%%%%%%%%%%%%%%%%%%%%%%%%%%%%%%%%%%%%%%%%%%%%%%%%%%%%%%%%%%
%%
%% START EDITING HERE!
%%
%%%%%%%%%%%%%%%%%%%%%%%%%%%%%%%%%%%%%%%%%%%%%%%%%%%%%%%%%%%%%%%%%%%%%%%%
\begin{titlepage}
\pubdate{\today}                    %fill in the date
%\pubnumber{XXX-XXXXX \\ YYY-YYYYYY} %preprint number(s)
\vfill
\Title{Multiple Peaks in the CMB}
\vfill
\Author{Alessandro ~Melchiorri$^{1}$}
\Address{$^{1}$ Denys Wilkinson Building,
University of Oxford, Keble Road, Oxford, OX1 3RH, UK.}
\vfill

\begin{Abstract}
Recent measurements of the Cosmic Microwave Background Anisotropy 
have provided evidence for the presence of oscillations in the
angular power spectrum. These oscillations are a wonderful
confirmation of the standard cosmological scenario and allow us to derive 
constraints on many cosmological, astrophysical and inflationary 
parameters. If the discovery is confirmed by future experiments,
opportunities may appear, for example, to constrain dark energy, 
variations in fundamental constants and neutrino physics.

\end{Abstract}

\vfill
\begin{Presented}
    COSMO-01 \\
    Rovaniemi, Finland, \\
    August 29 -- September 4, 2001
\end{Presented}
\vfill
\end{titlepage}
\def\thefootnote{\fnsymbol{footnote}}
\setcounter{footnote}{0}

%%%%%%%%%%%%%%%%%%%%%%%%%%%%%%%%%%%%%%%%%%%%%%%%%%%%%%%%%%%%%%%%%%%%%%%%
% The document starts here
%%%%%%%%%%%%%%%%%%%%%%%%%%%%%%%%%%%%%%%%%%%%%%%%%%%%%%%%%%%%%%%%%%%%%%%%

\section{Introduction}

In the last $2$ years important progress has been made in the
study of the Cosmic Microwave Background (CMB) Anisotropies.

With the TOCO$-97/98$ (\cite{torbet},\cite{miller}) 
and Boomerang-$97$ (\cite{mauskopf}) experiments a firm detection of
the first peak in the CMB anisotropy angular power spectrum has been
obtained. 
The presence of this peak is generally expected in
models of structure formation with a nearly-scale 
invariant spectrum of primordial perturbations 
like the one produced after inflation.
In the framework of adiabatic Cold Dark Matter (CDM) models, the
position, amplitude and width of this peak provide strong supporting 
evidence for the inflationary predictions of
a low curvature (flat) universe and a scale-invariant primordial 
spectrum (\cite{knox}, \cite{melchiorri}, \cite{tegb97}).

A first analysis of a small fraction of data from the
BOOMERANG $1998/1999$ Long Duration Ballooning (BOOM/LDB) campaign
(\cite{debe00}, \cite{lange}) and of observations from the MAXIMA experiment 
(\cite{hanany}, \cite{balbi}) 
further confirmed the presence of this feature at high 
significance. However, the finding
of a suppressed second peak in the CMBR anisotropy resulted in a rather
large value for the baryon density, $\Omega_b h^2 = 0.032^{+0.005}_{-0.004}$
at $68\%$ CL~\cite{th1}, while the experimental data on primordial $^4He$
and $D$ abundances, prefer smaller values, $\Omega_b h^2 = 0.020 {\pm}
0.002$ (\cite{burles}) (see also \cite{avelino},\cite{Lisi}).
Many authors addressed the issue of this tension between 
the determination of $\Omega_b h^2$ from CMBR data and 
Standard Big Bang Nucleosynthesis (SBBN) 
\cite{th4,peloso,th7,Hansen,Kaplin,dibari}.

The new experimental data from BOOMERANG (\cite{netterfield})
and DASI (\cite{halverson}) have refined the data at larger
multipole and now clearly suggest the presence of 
a second peak in the spectrum and a smaller value for the
baryonic fraction, in agreement with SBBN. 
Moreover, this result confirms the model
prediction of acoustic oscillations in the primeval plasma 
and shed new light on various cosmological and 
inflationary parameters (\cite{debe01}, \cite{wang}, \cite{pryke}).
The new results from MAXIMA (\cite{lee}, \cite{stompor}) 
are of lower precision, but are consistent with both DASI and 
BOOMERANG.

This paper is organized as follows: Section II is a 
brief introduction about why we expect oscillations in the
CMB spectrum. In Section III I will discuss the statistical significance
of the peaks measured by BOOMERANG, DASI and MAXIMA 
and their location and amplitude. In section IV I will review 
the implications for the cosmological parameters in the framework
of the standard CDM model
of structure formation. In section V I will discuss some 
non-standard aspect of parameter extraction.
Finally, in section VI, I will give my conclusions.

\section{The acoustic oscillations in the CMB anisotropy angular 
power spectrum}

\subsection{The power spectrum.}

The anisotropy with respect to the mean temperature $\Delta T=T-T_0$
 of the CMB sky in the direction $\bn$ measured at
time $t$ and from the position $\vec x$ can be expanded in 
spherical harmonics:

\be
{\De T\over T_0}(\bn,t,\vec x) = \sum_{\ell=1}^\infty\sum_{m=-\ell}^{m=\ell}
	 a_{\ell m}(t, \vec x) Y_{\ell m}(\bn)~,
\ee

If the fluctuations are Gaussian all the statistical information is
contained in the $2$-point correlation function. In the case of isotropic
fluctuations, this can be written as:

\be
\left \langle{\De T\over T_0}({\bn_1}){\De T\over T_0}({\bn_2})\right\rangle =
 {1\over 4\pi} \sum_\ell (2\ell+1)C_\ell P_\ell(\bn_1\cdot\bn_2)~.
\ee

where the average is an average over "all the possible universes"
i.e., by the ergodic theorem, over $\vec x$. The CMB power
spectrum $C_\ell$ are the ensemble average of the coefficients
$a_{\ell m}$,
\[ C_\ell = \langle|a_{\ell m}|^2\rangle ~. \]

Since it is impossible to measure ${\De T\over T_0}$ in every position in
the universe, we cannot do an ensemble average.
This introduces a fundamental limitation for the precision of a 
measurement (the cosmic variance) which is 
important especially for low multipoles. 
If the temperature fluctuations are Gaussian, the $C_{\ell}$ have 
a chi-square distribution with $2\ell+1$ degrees of freedom and 
the observed mean deviates from the ensemble average by 

\be 
{{\Delta C_\ell} \over C_\ell} = 
	\sqrt{2\over 2\ell + 1}~.  \label{2cv}
\ee

Moreover, in a real experiment, one never obtain complete sky 
coverage because of the limited amount of observational time 
(ground based and balloon borne
experiments) or because of galaxy foreground contamination 
(satellite experiments). 
All the telescopes also have to deal with the noise of the detectors 
and are obviously not sensitive to scales smaller than the 
angular resolution.
For a given an experiment, the accuracy of reconstruction of the 
power spectrum can be approximately given as \cite{knoxbeam}:

\be {{\Delta C_\ell} \over C_\ell} \simeq 
	\sqrt{2\over (2\ell + 1)f_{sky}} \left(1+
{{\sigma^2_{pixel}\Omega_{pixel}} \over C_\ell}
\exp[\ell^2\sigma^2_{beam}] \right)~.  \label{2cv2}
\ee

where $f_{sky}$ is the sky coverage, 
$\sigma_{beam}$ is the angular resolution, $\sigma_{pixel}$ the
experimental noise per pixel and $\Omega_{pixel}$ is
the area per pixel.

\subsection{Theoretical predictions. Inflation vs. Topological Defects.}

Acoustic oscillations in the CMB angular spectrum
 have been predicted since long time from simple assumptions about scale 
invariance and linear perturbation theory 
(\cite{Peeb1970}, \cite{SZ70}, \cite{wilson}, \cite{vittorio}, 
\cite{bondefstathiou}).
The physics of these oscillations and their dependence on 
the various cosmological parameters has been described in great detail
in many reviews (see e.g. \cite{review}, \cite{review2}, \cite{review3},
\cite{review4}, \cite{review5}, \cite{review6}). 
Here I will just outline the basic principles and spend a 
few words on models based on topological defects that 
{\bf do not} predict oscillations.

Since the CMB fluctuations are small, applying linear
perturbation theory to the Friedman metric is justified. 
For a cosmic fluid consisting of
radiation, massless neutrinos, baryons, cold dark
matter and a cosmological constant, we can the write down 
the linear perturbation equations (in Fourier space). 
In the most general case, for each wave vector $\bf k$ they are of
the form 
\be 
  {\bf D}X = {\cal S}~, \label{diff}
\ee
where $X$ is a vector containing all the random perturbation variables,
$\bf D$ is a deterministic linear first order differential operator and
${\cal S}$ is a random source term which consists of linear combinations 
of the energy momentum tensor of possible external sources 
of perturbations such as topological defects. 
More details can be found, {\em e.g.} in Ref.~\cite{DKM}, \cite{DKMrep}.

In general, for {\bf inflationary} perturbations ${\cal S}=0$ and the 
solutions are determined entirely by the random initial conditions, 
$X({\bf k},t_{\rm in})$.
Practically, scale-invariant perturbations are induced in the metric
only during the inflationary epoch by quantum fluctuations of the 
inflaton field. After that, no new perturbations are produced in 
the universe and everything evolves according to the linearized perturbation
equations.

For inflationary models  $X(\bk,t_{\rm in})$ is then a set of
Gaussian random variables and hence their statistical properties are
entirely determined by the spectra $\cal P$ (the Fourier transforms of the two
point  functions),
\be
	\langle X_i((t_{\rm in},\bk)\ X_j^*((t_{\rm in},\bk')\rangle
	\equiv {\cal P}_{ij}(\bk)\de(\bk-\bk')~. \label{ininf}
\ee  

Here the Dirac delta is a consequence of statistical homogeneity which
we want to assume for the random process leading to the initial
perturbations. In general, for density inflationary perturbations, 
${\cal P} \sim k^{n_S}$ where 
the spectral index $n_S$ is equal to $1$ for scale-invariance. 

Let $A_i(k,t)$ be the solution with initial condition   
$X_j(k,t_{\rm in})$. 
The spectra of the solution with initial 'spectrum' given by
Eq.~(\ref{ininf}) is then today just
\[
 \langle X_i((t_0,\bk)\ X_j^*((t_0,\bk')\rangle =A_i(k,t_0)A_j^*(k,t_0)
 {\cal P}_{ij}(k)\de(\bk-\bk')~. 
\]

Therefore, if $A_i$ is oscillating, {\em e.g.}
as a function of $kt$ because of some physical process
between $t_{\rm in}$ and $t_0$, so will $ \langle |X_i|^2\rangle$. 

This is exactly our case: on sub-horizon scales, 
prior to recombination, photons and baryons form a tightly 
coupled fluid that performs acoustic 
oscillations driven by the gravitational potential. 
These acoustic oscillations define a structure of peaks in
the CMB angular power spectrum that is measured today.

Let us, for example, consider the density fluctuations $\delta$ in the
baryon-photon fluid during the radiation dominated epoch.
In this case we have $w=p/\rho\simeq 1/3$ equal to the
adiabatic sound speed $c_s^2={\dot p/ \dot \rho}\simeq 1/3$
and $\delta$ follows the wave equation:

\be
\ddot{\delta} +k^2{\delta\over 3}= {k^2 \over 3} (\Phi-\Psi) \label{ac2}
\ee

where $\Phi$ and $\Psi$ are the Bardeen potentials ~\cite{Bardeen},
and the dots are derivatives respect to the time $\eta$. 
In our case $\Phi \simeq -\Psi \simeq const$.
On very large, super-horizon scales, $k\eta\ll 1$, 
$\delta$ remains constant. Once $k\eta > 1$ $\delta$ 
begins to oscillate like an acoustic wave. 

If adiabatic perturbations have been created during an early inflationary 
epoch, they all start oscillating in phase. At the moment of 
recombination, when the photons become free and the acoustic oscillations 
stop, the perturbations of a given wave length thus all have the same 
phase. As each given wave length is projected to a fixed angular 
scale on the sky, this leads to a characteristic structure of peaks 
in the CMB power spectrum. 

However, if the source term $\SSS$ does not vanish, 
the randomness of the source term enters at all times 
and the situation is different.

In this case, Equation~(\ref{diff}) can be solved by
means of a Green's function, ${\cal G}(t,t')$, in the form
\be
X_j(t_0,\bk) =\int_{t_{in}}^{t_0}\! dt{\cal G}_{jl}(t_0,t,\bk){\SSS}_l(t,\bk)~.
\label{Gsol}
\ee
Power spectra or, more generally, quadratic  expectation
values of the form
$\langle X_j(t_0,\bk)X_l^*(t_0,\bk)\rangle $ are then given by

\be
\langle X_j(t_0,\bk)X_l^*(t_0,\bk)\rangle =
 \int_{t_{in}}^{t_0}\! dt \int_{t_{in}}^{t_0} \! dt' {\cal G}_{jm}(t_0,t,\bk)
  {\cal G}^*_{ln}(t_0,t',\bk)
 \langle{\SSS}_m(t,\bk){\SSS}_n^*(t',\bk)\rangle~. \label{power}
\ee

The only information about the source random variable which we really
need in order to compute power spectra are therefore the unequal time
 two point correlators (see e.g. ~\cite{ACFM,PST,Aetal,DS,DKM}

\be
\langle{\SSS}_m(t,\bk){\SSS}_n^*(t',\bk)\rangle~. \label{2point}
\ee

In the case of topological defects or more general 
non homogeneous distributions of matter, $\SSS$ is given by a function 
quadratic in the defect field, which itself obeys highly non-linear 
evolution equations. The non-linearity of the time evolution
of the source term has several important consequences.
Even though time evolution is deterministic, different Fourier 
modes mix due to non-linearity, and the
randomness in one mode 'sweeps' into the other modes. 

Therefore, fluctuations of a given wave number $k$ are in
general not in phase, and the distinctive series of acoustic peaks
present in inflationary models is blurred into one 'broad hump'. 

In Figure 1 we plot the $2$ (very different) 
theoretical predictions one for an inflationary model and one for a 
model based on numerical simulations of global textures taken from
\cite{DKM}. Global textures can be considered a good representative 
for the models with non linear sources. 
The cosmological parameters are assumed to be the same.

As we can see, when non-linearities are present as in the textures case, 
decoherence dominates and the oscillations in the CMB spectrum are severely 
damped. 
Also plotted in the figure are the recent data from the BOOMERanG and
DASI experiments. It is clear from the picture that while the inflationary
scenario is in good agreement with the data, the texture model is
ruled out at high significance.

This is an important result: structure formation theories based
on non-linear external sources fail to match the current data, and
defects as only responsible mechanism for structure formation 
can reasonably be ruled out.

However, before going to parameter extraction, is important to asses
how well the current data are in agreement with the underlying theoretical
model and with the presence of acoustic oscillations.

\section{Are there acoustic oscillations in the CMB power spectrum ?}

\begin{figure}[htb]
\begin{center}
\includegraphics[angle=-90,scale=0.45]{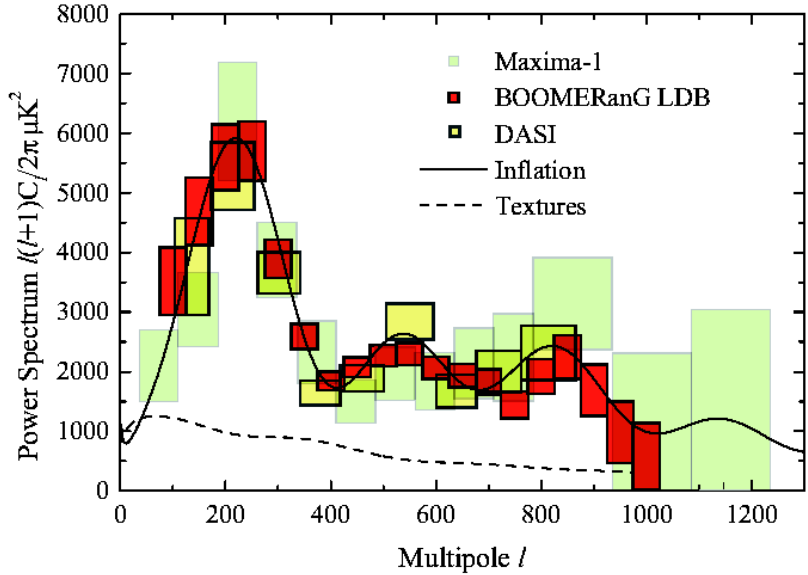}
\end{center}
\caption{BOOMERanG, DASI and MAXIMA data togheter with an inflationary model
and a global textures model.}
\label{fig1}
\end{figure} 

On April 30th 2001, at the same time, $3$
different teams,  Boomerang \cite{netterfield}, DASI \cite{halverson} 
and MAXIMA \cite{lee} reported a detection of multiple features in 
the CMB angular power spectrum. 
Before discussing the statistical significance of these features,
let me briefly review the various experiments.

\subsection{The BOOMERanG experiment.}

BOOMERanG is a scanning balloon experiment aimed at producing accurate 
and high signal/noise maps of the CMB sky and constraining the power
spectrum in the $50 < \ell < 1000$ range. 
The BOOMERanG experiment has been described in 
\cite{piacentini} and \cite{debe99}. 
All the relevant informations about the collaboration can be found
in the 'official' websites: {\it \bf http://oberon.roma1.infn.it/boom} and 
{\it \bf http://www.physics.ucsb.edu/\~boomerang/}.

The BOOMERanG group carried out a long duration flight 
(December 1998/ January 1999) called the Antarctica or
LDB flight. Before this, there was a 'test flight' on North
America from which the first power spectrum results were
released (\cite{mauskopf}, \cite{melchiorri}, \cite{piacentini}).
From the test flight a $\sim 4000$ $16'$ pixel map 
at $150 GHz$ produced a firm detection of 
a first peak in the CMB angular power spectrum.

For the antarctica flight, coverage of $4$ frequencies with $16$
bolometers in total were available. 
BOOMERanG LDB measured $8$ pixels in the sky simultaneously . 
Four pixels feature multiband photometers ($150$, $240$ and
$410$ GHz), two pixels have single-mode, diffraction limited
detectors at $150$ GHz and two pixels have single-mode,
diffraction limited detectors at $90$ GHz.
The NEP of these detectors is below $200 \mu K_{CMB}\sqrt{s}$ at 
$90$, $150$ and $240$ GHz and the angular resolution ranges
from $12$ to $18$ arcminFWHM.

The istrument was flown aboard a stratospheric balloon at
$38 Km$ of altitude to avoid the bulk of atmospheric emission
and noise.
During a long duration balloon flight of $\sim 11$ days 
carried out by NASA-NSBF around Antarctica in $1999$, 
BOOMERanG mapped $\sim 1800$ square degrees in a region of the 
sky with minimal contamination from the galaxy.

The most recent analysis of the BOOMERanG data has been presented 
in \cite{netterfield}. The observations taken from $4$ 
detectors at $150$ GHz in a dust-free ellipsoid central region 
of the map ($1.8 \%$ of the sky) have been analyzed using the
methods of (\cite{borrill}, \cite{hivon}, \cite{prunet}).
The gain calibration are obtained from observations of the CMB dipole.

The CMB angular power spectrum, estimated in $19$ bands
centered between $\ell=50$ to $\ell=1000$ is shown in Figure 1.
The error bars on the $y$ axis are correlated at about $\sim 10 \%$.
A first peak is clearly evident at $\ell \sim 200$ and 
$2$ subsequent peaks can be see in the figure.
Not shown in the figure is an additional $10 \%$
calibration error (in $\Delta T$) and the uncertainty in the beam 
size ($12.9' \pm 1.4'$).

Calibration error does not affect the shape of the power spectrum,
producing only an overall shift in amplitude.

On the contrary, since the beam resolution affects the 
$C_\ell$ spectrum according to the Eq. (\ref{2cv2}), a small uncertainty 
$\Delta \sigma^2_{beam}=\sigma^2_{beam}-(\sigma_{beam}')^2$ 
in the telescope beam produce a correlated $\ell$-dependent 
'calibration error' of $\sim (1+\ell^2\Delta\sigma^2_{beam})C_\ell$
(\cite{bridle})

The beam uncertainty can change the relative amplitude
of the peaks, but cannot introduce features in the spectrum.

\subsection{The DASI experiment.}

The DASI experiment is a ground based compact interferometer 
constructed specifically for observations of the CMB.
A description of the instrument can be found in 
\cite{halverson} and \cite{leitch}.
and all the relevant information about the team can be
obtained from the DASI website:{\it \bf http://astro.uchicago.edu/dasi/}.

The specific advantage of interferometers is in reducing the
effects of atmospheric emission \cite{lay}.
DASI is composed of $13$ element interferometers
with correlator operating from $26$ to $36$ GHz.
The baseline of DASI cover angular scales from $15'$ to
$1.4^o$.

Interferometry is a technique that differs in many
fundamental ways from those used by BOOMERanG and other
map-making CMB experiments.
Interferometers directly sample the Fourier
transform of the sky brightness distribution and 
the CMB power spectrum can be computed without
going through the map making process.
In this sense, the DASI result provides a real
independent observation of the CMB angular spectrum.

The most recent analysis of the DASI data has been 
presented in \cite{halverson}. 
The observations have been taken over $97$ 
days from the South-Pole during the austral summer
at frequencies between  $26$ and $36$ GHz.
The calibration was obtained using bright astronomical
sources.

The CMB angular power spectrum estimated in $9$ bands
between $\ell=100$ to $\ell=900$ is also shown in Figure 1.
There is a $\sim 20 \%$ correlation between the data points.
Not shown in the figure is an $\sim 8\%$ calibration error,
while the beam error is negligible.
The DASI team found no evidence for foregrounds other than
point sources (which are the dominant foregrounds at those
frequencies (see e.g. \cite{efstteg}, \cite{tegfor})).
Nearly $30$ point sources have been detected in the DASI
data while a statistical correction has been made for residual
point sources that were too faint to be detected.

\subsection{The MAXIMA experiment.}

MAXIMA-I is another balloon experiment, similar in many aspects
to BOOMERang but not long-duration.
A description of the instrument can be found in \cite{lee}
and all the relevant informations about the team can be
obtained from the MAXIMA website:
{\it \bf http://cosmology.berkeley.edu/group/cmb/}.       
In the latest analysis (\cite{lee}) the data from 
$3$, $150$, GHz very sensitive bolometers has been analyzed in order to
produce a $3'$ pixelized map of about $10$ by $10$ degrees. 
The previous analysis of about the same data\footnote{The $240$ GHz channel 
has been excluded because it did not pass consistency
tests above $\ell=785$.} based on a $5'$ pixelization
(\cite{hanany}) has therefore been extended to $\ell=1235$.
The map-making method used by the MAXIMA team is extensively discussed
in \cite{stompor2}. The data are calibrated using the CMB dipole.

The MAXIMA-I datapoints are also shown in Figure 1.
The error bars are correlated at level of
$\sim 10 \%$. The $\sim 4 \%$ calibration error is not plotted in 
the figure. The beam/pointing  errors are of order of $\sim 10 \%$ at
$\ell =1000$ (see \cite{lee}).

\subsection{Features in the CMB power spectrum.}

\begin{figure}[htb]
\begin{center}
\includegraphics[scale=0.35]{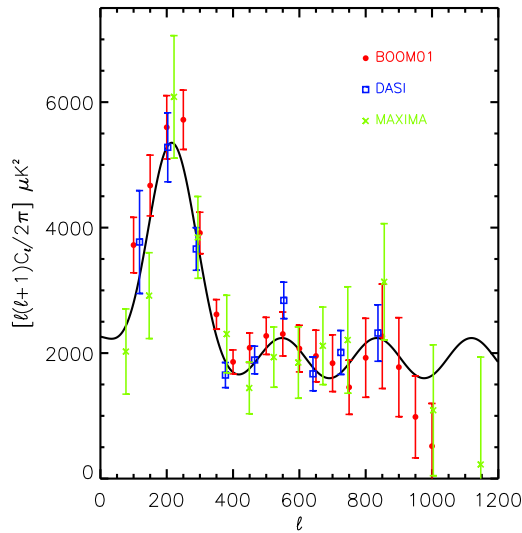}
\end{center}
\caption{Phenomenological fit to the BOOMERanG, DASI and Maxima data.
Picture taken from \cite{douspis}}
\label{fig2}
\end{figure}

Before going on to parameter extraction, 
it is important to adopt a phenomenological approach and try to quantify 
how well the present data provide evidence for multiple and coherent
oscillations. 
Fits to the CMB data with phenomenological functions have been 
already extensively used in the past (see e.g. \cite{graca}, 
\cite{scott}, \cite{page}).
More recently, similar analyses have been carried out,
using parabolas (\cite{debe01}, \cite{bohdan}) 
or more elaborate oscillating functions with a well defined
frequency and phase (\cite{douspis}).

Since the first peak is evident, the statistical significance
of the secondary oscillations is now of greater interest.
In \cite{debe01} the BOOMERanG data bins centered 
at $450 < \ell < 1000$ were analyzed.
Using a Bayesian approach, a
 linear fit $C_\ell^T = C_A + C_B \ell$ is rejected at near 
$2 \sigma$ confidence level.
Also in \cite{debe01}, using a parabolic fit to the data, 
interleaved peaks and dips were found at $\ell =$ $215 \pm 11$, $431 \pm 10$,
$522 \pm 27$, $736 \pm 21$ and $837 \pm 15$ with 
amplitudes of the features $5760^{+344}_{-324}$, 
$1890^{+196}_{-178}$, $2290^{+330}_{-290}$,
$1640^{+500}_{-380}$, and $2210^{+900}_{-640}$ $\mu K^2$, correspondingly. 
The reported significance of the detection is $1.7\,\sigma$ for 
the second peak and dip, and $2.2\,\sigma$ for the third peak. 

The evidence for oscillations in the MAXIMA data 
has been carefully studied in \cite{stompor}.
While there is no evidence for a second peak, the power 
spectrum shows excess power at $\ell \sim 860$ over the average 
level of power at $411 \le\ell \le 785$ on the $95 \%$ confidence level.
Such a feature is consistent with the presence of a third acoustic peak.
         
In \cite{bohdan} the BOOMERanG, DASI and MAXIMA data were included in
a similar analysis. Both DASI and MAXIMA confirmed the main features
of the Boomerang CMB power spectrum: a dominant first acoustic peak
at $\ell \sim 200$, DASI shows a second peak at $\ell \sim 540$
and MAXIMA-I exhibits mainly a 'third peak' at $\ell \sim 840$.

Finally and more recently, in \cite{douspis} a different analysis 
was made, based on a function that smoothly interpolates between a spectrum 
with no oscillations and one with oscillations.
Again, within the context of this different phenomenological model,
a $2 \sigma$ presence for secondary oscillations was found.
In Figure 2 the best phenomenological fit to the
data from \cite{douspis} is reported. 
As we can see, the oscillations are clearly present in the 
BOOMERanG and DASI data and are compatible with the MAXIMA data.

\section{Consequences for Cosmology}

Since the observations are, at the very least, 
compatible with acoustic oscillations, we can make the assumption
of inflationary perturbations and undertake a parameter estimation.

Constraining the parameters of the model with the
present CMB data can be regarded as a further test for
the consistency of the scenario, since one can then compare 
the results with those obtained by independent methods 
and/or under different theoretical assumptions.

In principle, the CDM scenario of structure formation based on adiabatic
primordial fluctuations can depend on more than $11$ parameters.

However for a first analysis, it is useful to restrict ourselves to 
just $5$ parameters: the tilt of primordial spectrum of perturbations 
$n_S$, the optical depth of the universe $\tau_c$,
the density in baryons and dark matter
$\omega_b=\Omega_bh^2$ and $\omega_{dm}=\Omega_{dm}h^2$ and 
the shift parameter ${\cal R}$ which is related to the geometry of the universe
through (see \cite{efsbond}, \cite{melou}):

\be
{\cal R}=2 \sqrt{|\Omega_k| / \Omega_m} / \chi(y)
\ee

where $\Omega_m=\Omega_b+\Omega_{dm}$, $\Omega_k=1-\Omega_m-\Omega_{\Lambda}$,
the function $\chi(y)$ is $y$, $\sin(y)$ or $\sinh(y)$ for flat, closed and
open universes respectively and

\be
y=\sqrt{|\Omega_k|}\int_0^{z_{dec}}
{[\Omega_m(1+z)^3+\Omega_k(1+z)^2+\Omega_{\Lambda}]^{-1/2} dz}.
\ee

\begin{figure}[htb]
\begin{center}
\includegraphics[scale=0.38]{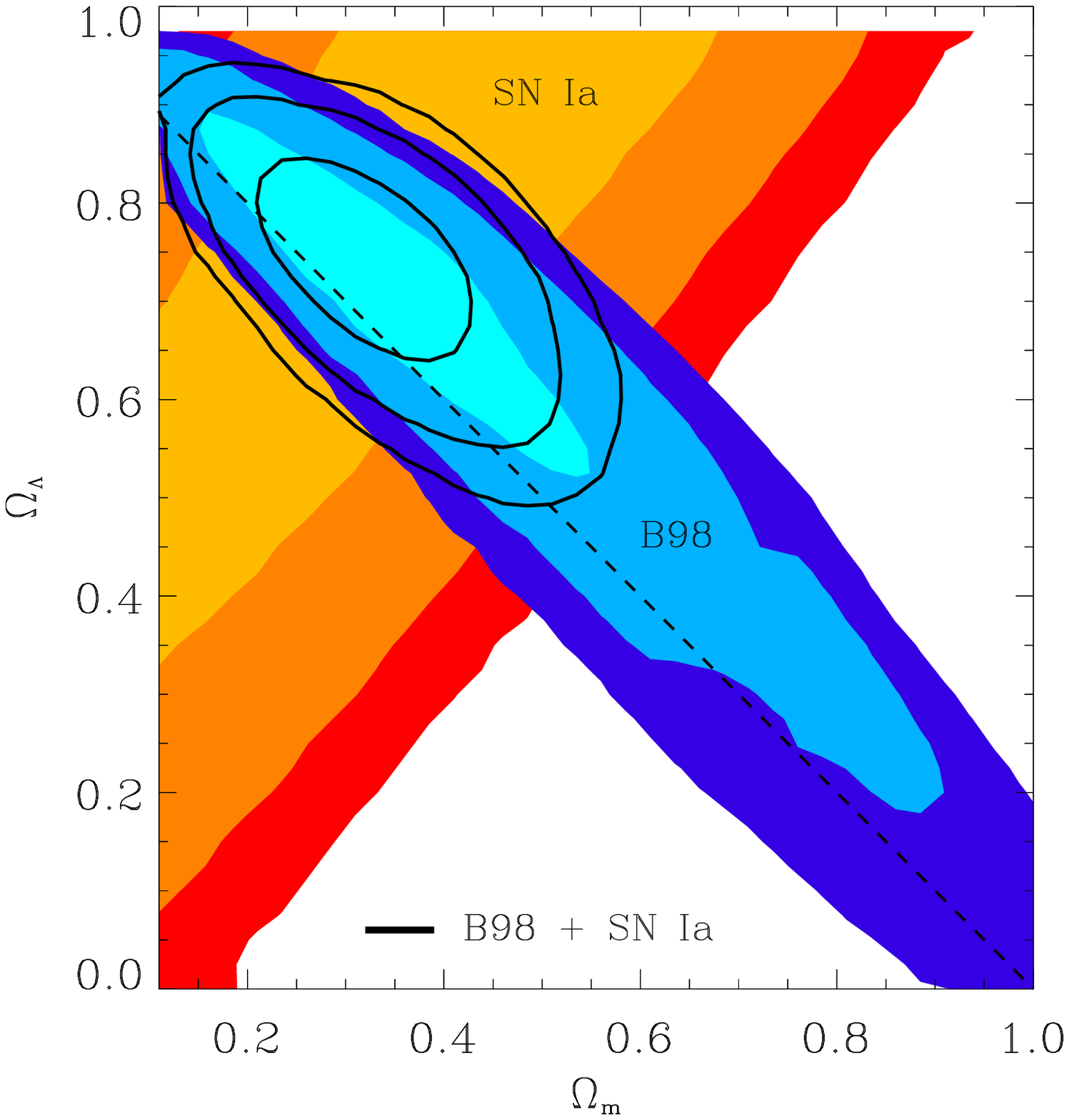}
\includegraphics[scale=0.38]{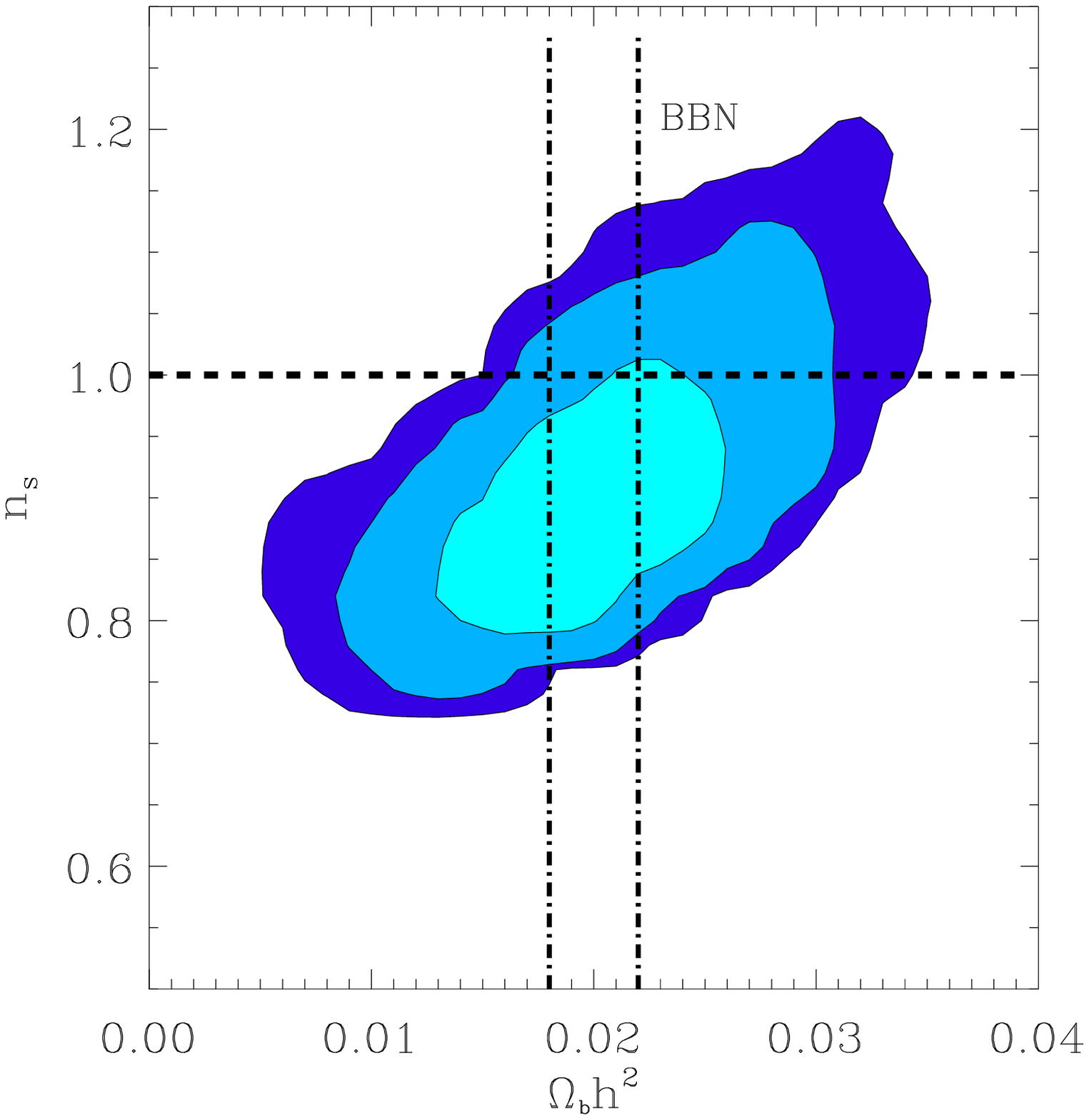}
\end{center}
\caption{Confidence contours in the $\Omega_M - \Omega_{\Lambda}$ 
and $\Omega_bh^2-n_S$ planes. Picture taken from \cite{debe01}.}
\label{fig3}
\end{figure}

The restriction of the analysis to only $5$ parameters can be justified 
in several way: First, a reasonable fit to the
data can be obtained with no additional parameters. 
Second, constraints on these parameters provide a 
test of the scenario: the value of the baryon density 
$\Omega_bh^2$ can  be compared with the values obtained 
from Big Bang Nucleosynthesis; $\Omega_{dm}h^2$ must be significantly 
different from zero since a purely baryonic model doesn't match the 
observed galaxy clustering; $\Omega=1$ and $n_S \sim 1$ are general 
predictions of the inflationary scenario.

Let us consider the results from BOOMERanG, DASI and MAXIMA separately,
joint analysis have been reported in, for example, 
(\cite{wang}) and (\cite{douspis}).

In Fig. 3 we plot the likelihood contours in the 
$\Omega_{M}-\Omega_{\Lambda}$ and $\Omega_bh^2-n_S$ planes
from the BOOMERanG experiment as reported in \cite{debe01}.
Since the quantity ${\cal R}$ depends on $\Omega_{\Lambda}$ and 
$\Omega_{M}$ the CMB constraints on this parameter can be 
plotted on this plane.
As we can see from the left panel in the figure 
the data strongly suggest a flat universe
(i.e. $\Omega=\Omega_M+\Omega_{\Lambda}=1$). From the 
latest BOOMERanG data one obtains $\Omega=1.02\pm0.06$ (\cite{netterfield}).

The inclusion of complementary datasets in the analysis
breaks the angular diameter distance degeneracy in $\cal R$ 
and provides evidence
for a cosmological constant at high significance.
Adding the Hubble Space Telescope constraint on the Hubble
constant $h=0.72 \pm 0.08$ (\cite{freedman}, 
information from galaxy clustering and
from luminosity distance of type Ia supernovae 
gives (\cite{netterfield}) $\Omega_{\Lambda}=0.62_{-0.18}^{+0.10}$, 
$\Omega_{\Lambda}=0.55_{-0.09}^{+0.09}$ and 
$\Omega_{\Lambda}=0.73_{-0.07}^{+0.10}$ respectively.

Also interesting is the plot of the likelihood contours in the 
$\Omega_bh^2-n_S$ plane. 
These $2$ parameters are crucial in the determination of the relative
amplitude of the peaks and the power on subdegree angular scales.
Namely, increasing the baryon density, increases the difference between the
first and second peak. Decreasing the scalar spectral index has the same
effect since we are removing power from small scales and adding it to 
large angular scales.
Therefore, some sort of degeneracy exists among these parameters that 
can however be broken by measuring the power around the $3$rd peak.
The presence of the degeneracy is exemplified by the elongation
of the likelihood contours along the $\omega_b-n_S$ direction.
The present data, however, already provide enough information about
the power on the $3$rd peak and the degeneracy is partially broken.

The most important result from the right panel of Figure 3 
is that the present BOOMERanG data is in beautiful agreement with 
{\it both} a nearly scale invariant
spectrum of primordial fluctuations, as predicted by inflation, and
the value for the baryon density $\omega_b =0.020\pm0.002$ predicted
by Standard Big Bang Nucleosynthesis (see e.g. \cite{burles}).

\begin{figure}[htb]
\begin{center}
\includegraphics[scale=0.38]{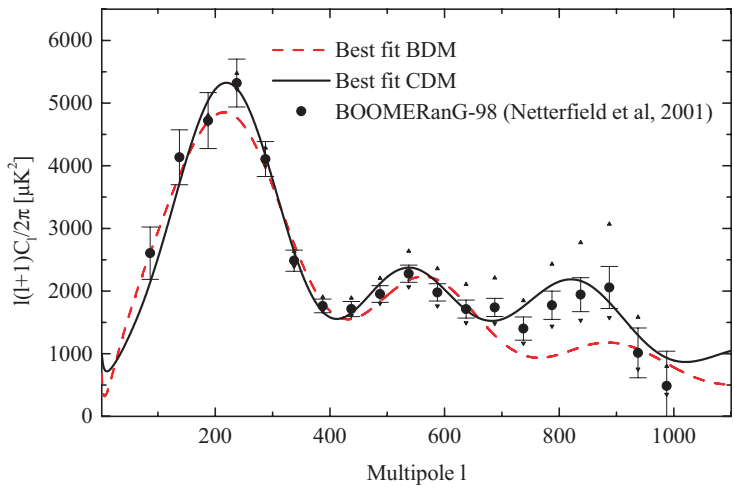}
\end{center}
\caption{CDM model vs purely baryonic dark matter models. Picture
taken from \cite{bmp}.}
\label{fig5}
\end{figure}

An increase in the optical depth $\tau_c$ after recombination 
by reionization (see e.g. \cite{haiman} for a review) or by some more
exotic mechanism damps the amplitude of the CMB peaks.
Even if degeneracies with other parameters such as $n_S$ are present
(see e.g. \cite{debe97}) the BOOMERanG data provides the 
upper bound $\tau_c < 0.3$.

The amount of non-baryonic dark matter is also 
constrained by the CMB data with $\Omega_{dm}h^2=0.13 \pm 0.04$ 
at $68 \%$ c.l. (\cite{netterfield}).
The presence of power around the third peak is crucial in this sense,
since it cannot be easily accommodated in models based on just baryonic
matter (see e.g. \cite{bmp}, \cite{lmg}, \cite{mcgaugh} 
and references therein).
In Fig.4 we plot the BOOMERanG data
with the best fit purely baryonic (BDM) and CDM model (picture taken from
\cite{bmp}). As we can see, BDM models fail to reproduce the observed power
at $\ell \ge 700$.

Furthermore, under the assumption of flatness, we can derive important
constraints on the age of the universe $t_0$ given by:

\begin{equation}
t_0=9.8 Gy \int_0^1{{a da \over
{[\omega_ma+\omega_{\Lambda}a^4]^{1/2}}}}
\end{equation}

\begin{figure}[htb]
\begin{center}
\includegraphics[scale=0.35]{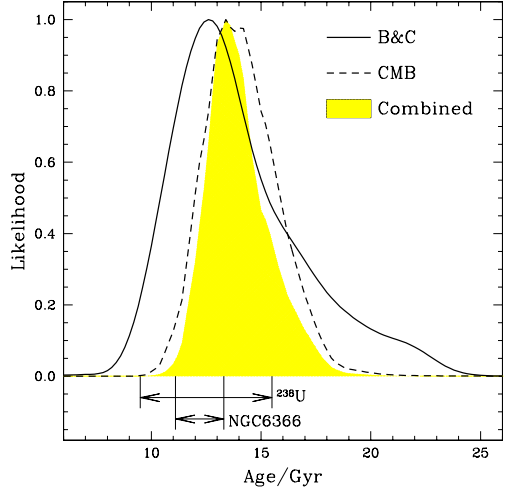}
\end{center}
\caption{Age constraint from CMB compared with independent estimates.
Picture taken from \cite{iggy}.}
\label{fig6}
\end{figure}

In Figure 5 we compare the BOOMERanG constraint on age with 
other independent results obtained from 
stellar populations in bright ellipticals (\cite{iggy}),
 $^{238}$U age-measurement of an old halo star in our 
galaxy (\cite{cayrel}) and age the of the oldest halo globular cluster 
in the sample of Salaris \& Weiss (\cite{salaris}). 
As we can see all four methods give completely consistent results,
and enable us to set rigorous bounds on the maximum and minimum ages
that are allowed for the universe, $t_0=14 \pm 1$ GYrs (\cite{iggy},
\cite{netterfield},\cite{knoxage}).

The results from the DASI experiment have been extensively
reported in \cite{pryke} and are perfectly consistent with the
BOOMERanG results.
Pryke et al. report $\Omega=1.04 \pm 0.06$, $n_s=1.01^{0.08}_{0.06}$, 
$\Omega_bh^2=0.022^{0.004}_{0.003}$ and $\Omega_{dm}h^2=0.14 \pm 0.04$. 

The MAXIMA team reported similar compatible constraints in 
\cite{stompor}: $\Omega=0.9{+0.18\atop-0.16}$ and
$\Omega_b h^2=0.033{\pm 0.13}$ at $2 \sigma$ c.l..
However the MAXIMA data is not good enough to put strong 
constrains on the spectral index $n_S$ and the optical depth 
$\tau_c$ because of the degeneracy between the $2$ parameters.

\section{Non-Standard Aspects of Parameter extraction}

Even if the present CMB observations can be fitted with just $5$ 
parameters it is interesting to extend the analysis to other 
parameters allowed by the theory.
Here I will just summarize a few of them and discuss
 how well we can constrain them and what the effects 
on the results obtained in the previous section would be.

\subsection{Gravity Waves}

The metric perturbations created during inflation belong to two types:
{\it scalar} perturbations, which couple to the stress-energy of 
matter in the universe and form the ``seeds'' for structure formation 
and {\it tensor} perturbations, also known as 
gravitational wave perturbations.
Both scalar and tensor perturbations contribute to CMB anisotropy.
In the recent CMB analysis by the BOOMERanG and DASI collaborations, 
the tensor modes have been neglected, 
even though a sizable background of gravity waves 
is expected in most of the inflationary scenarios. 
Furthermore, in the simplest models,
a detection of the GW background 
can provide information on the second derivative
of the inflaton potential and shed light on the physics at
$\sim 10^{16} Gev$ (see e.g. \cite{hoffman}).

The shape of the $C^T_{\ell}$ spectrum from tensor modes is drastically
different from the one expected from scalar fluctuations,
affecting only large angular scales (see e.g. \cite{crittenden}). 
The effect of including tensor modes is similar to 
just a rescaling of the degree-scale $COBE$ normalization and/or 
a removal of the corresponding data points from the analysis.

This further increases the degeneracies among cosmological
parameters, affecting mainly the estimates of the baryon and 
cold dark matter densities and the scalar spectral index $n_S$
(\cite{melk99},\cite{kmr}, \cite{wang}, \cite{efstathiougw}).

The amplitude of the GW background is therefore weakly constrained
by the CMB data alone, however, when information from BBN, local
cluster abundance and galaxy clustering are included, an upper limit
of about $r = C_2^T/C_2^S < 0.5$ is obtained. 

\subsection{Scale-dependence of the spectral index.}

The possibility of a scale dependence of the
scalar spectral index, $n_S(k)$, has been considered
in various works (see e.g. \cite{kosowsky}, \cite{copeland}, 
\cite{lythcovi}, \cite{doste}).
Even though this dependence is considered to 
have small effects on CMB scales in most of the slow-roll inflationary models, 
it is worthwhile to see if any useful constraint can be obtained.
Allowing the power spectrum to bend erases the ability 
of the CMB data to measure the tensor to scalar perturbation ratio
and enlarge the uncertainties on many cosmological parameters.

Recently, Covi and Lyth (\cite{covi}) investigated the
two-parameter scale-dependent spectral index
predicted by running-mass inflation models, and 
found that present CMB data allow for a significant scale-dependence of $n_S$.
In Hannestad et al. (\cite{hhv}, \cite{hhvh}) 
the case of a running spectral index has been studied, 
expanding the power spectrum $P(k)$ to second order in $ln(k)$. 
Again, their result indicates that a bend in the spectrum is
consistent with the CMB data.

Furthermore, phase transitions associated with spontaneous 
symmetry breaking during the inflationary era could result
in the breaking of the scale-invariance of the primordial density 
perturbation.
In \cite{sarkar}, \cite{louise} and \cite{ywang} 
the possibility of having step or bump-like features in the spectrum 
has also been considered.

While much of this work was motivated by the tension between the initial
release of the data and the baryonic abundance value from BBN, 
a sizable feature in the spectrum is still compatible with the latest
CMB data (\cite{elgaroy}).

\subsection{Quintessence}

The discovery that the universe's evolution may be dominated by
an effective cosmological constant \cite{super1}
is one of the most remarkable cosmological findings of recent years.
One candidate that could possibly explain the observations is a
dynamical scalar ``quintessence'' field. One of the strongest aspects of
quintessence theories is that they go some way towards explaining the
fine-tuning problem, that is why the energy density producing the acceleration
is $\sim 10^{-120}M_{pl}^{4}$. A vast range of ``tracker'' (see for
example \cite{quint,brax}) and ``scaling'' (for example
\cite{wett}, \cite{ferjoy}) quintessence models exist which approach
attractor solutions, giving the required energy density, independent
of initial conditions.
The common characteristic of quintessence
models is that their equations of state, $w_{Q}=p/\rho$, vary with time
while a cosmological constant remains fixed at
$w_{Q=\Lambda}=-1$ (see e.g. \cite{blud}). 
Observationally distinguishing a time variation in
the equation of state or finding $w_Q$ different from $-1$ will
therefore be a success for the quintessential scenario.
Quintessence can also affect the CMB by acting as an additional 
energy component with a characteristic viscosity.
However any early-universe imprint of quintessence 
is strongly constrained by Big Bang Nucleosynthesis
with $\Omega_Q(MeV) < 0.045$ at $2\sigma$ for temperatures near 
$T \sim 1Mev$ (\cite{bhm}, \cite{mathews}).

In \cite{rachel} we have combined the latest observations of the 
CMB anisotropies and the information from Large
Scale Structure (LSS) with the luminosity distance of high
redshift supernovae (SN-Ia) to put constraints on the dark energy
equation of state parameterized by a redshift independent
quintessence-field pressure-to-density ratio $w_Q$.                  
          
The importance of combining different data sets in order to obtain
reliable constraints on $w_Q$ has been stressed by
many authors (see e.g. \cite{PTW}, \cite{hugen},\cite{jochen}),
since each dataset suffers from degeneracies between the various
cosmological parameters and $w_Q$ . Even if one restricts consideration
 to flat universes and to a value of $w_Q$ constant
in time then the SN-Ia luminosity distance and position of the
first CMB peak are highly degenerate in $w_Q$ and $\Omega_Q$,
the energy density in quintessence.                              

Varying $w_Q$ on the angular power spectrum of the
CMB anisotropies has just $2$ effects.
First, since the inclusion of quintessence
changes the overall content of matter and energy, the angular
diameter distance of the acoustic horizon size at recombination 
will be altered.

In flat models (i.e. where the energy density in matter is 
$\Omega_M=1-\Omega_Q$), this creates a shift in the
peaks' positions in the angular spectrum by
\be
{\cal R}^{-1}={1 \over 2} \sqrt{(1-\Omega_Q)}\int_0^{z_{dec}}
[(1-\Omega_Q)(1+z)^3+\Omega_{Q}(1+z)^{3(1+w_Q)}]^{-1/2} dz.
\ee

It is important to note that the effect is completely degenerate
in the interplay between $w_Q$ and $\Omega_Q$.
Furthermore, it does not add any new
features beyond those produced by the presence of a
cosmological constant \cite{efsbond}, and it is not particularly sensitive
to further time dependencies of $w_Q$.

Second, the time-varying Newtonian potential after decoupling will
produce anisotropies at large angular scales through the Integrated
Sachs-Wolfe (ISW) effect. The curve in the CMB angular spectrum
on large angular scales depends not only on the value of
$w_Q$, but also on its variation with redshift.
This effect, however, will be difficult to
disentangle from the same effect generated by a cosmological constant,
especially in view of the affect of cosmic variance and/or
gravity waves on the large scale anisotropies.

In order to emphasize the importance of degeneracies among
all of these parameters while analyzing the CMB data, we plot
in Figure $6$ some degenerate spectra, obtained by keeping
the physical density in matter $\Omega_Mh^2$, the physical density
in baryons $\Omega_bh^2$ and
${\cal R}$ fixed. As we can see from the plot, models degenerate
in $w_Q$ can easily be constructed.
However the combination of CMB data with other
different datasets can break the mentioned degeneracies.                     

\begin{figure}
\begin{center}
%\epsfxsize=7.2cm
%\epsfysize=6.2cm
%\epsffile{fig1.eps}
\includegraphics[scale=0.38]{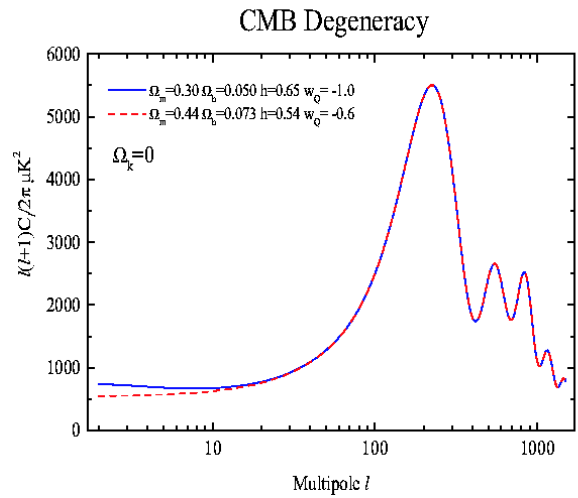}
\end{center}
\caption{CMB power spectra and the angular diameter distance
degeneracy. The models are computed assuming flatness,
$\Omega_k=1-\Omega_M-\Omega_Q=0$). The Integrated Sachs Wolfe effect on large 
angular scale slightly breaks the degeneracy. The degeneracy can be broken 
with a strong prior on $h$. Picture taken from \cite{rachel}.}
\label{fig7}
\end{figure}

\begin{figure}
\begin{center}
%\epsfxsize=6.2cm
%\epsfysize=7.2cm
%\epsffile{totrach.eps}
\includegraphics[scale=0.38]{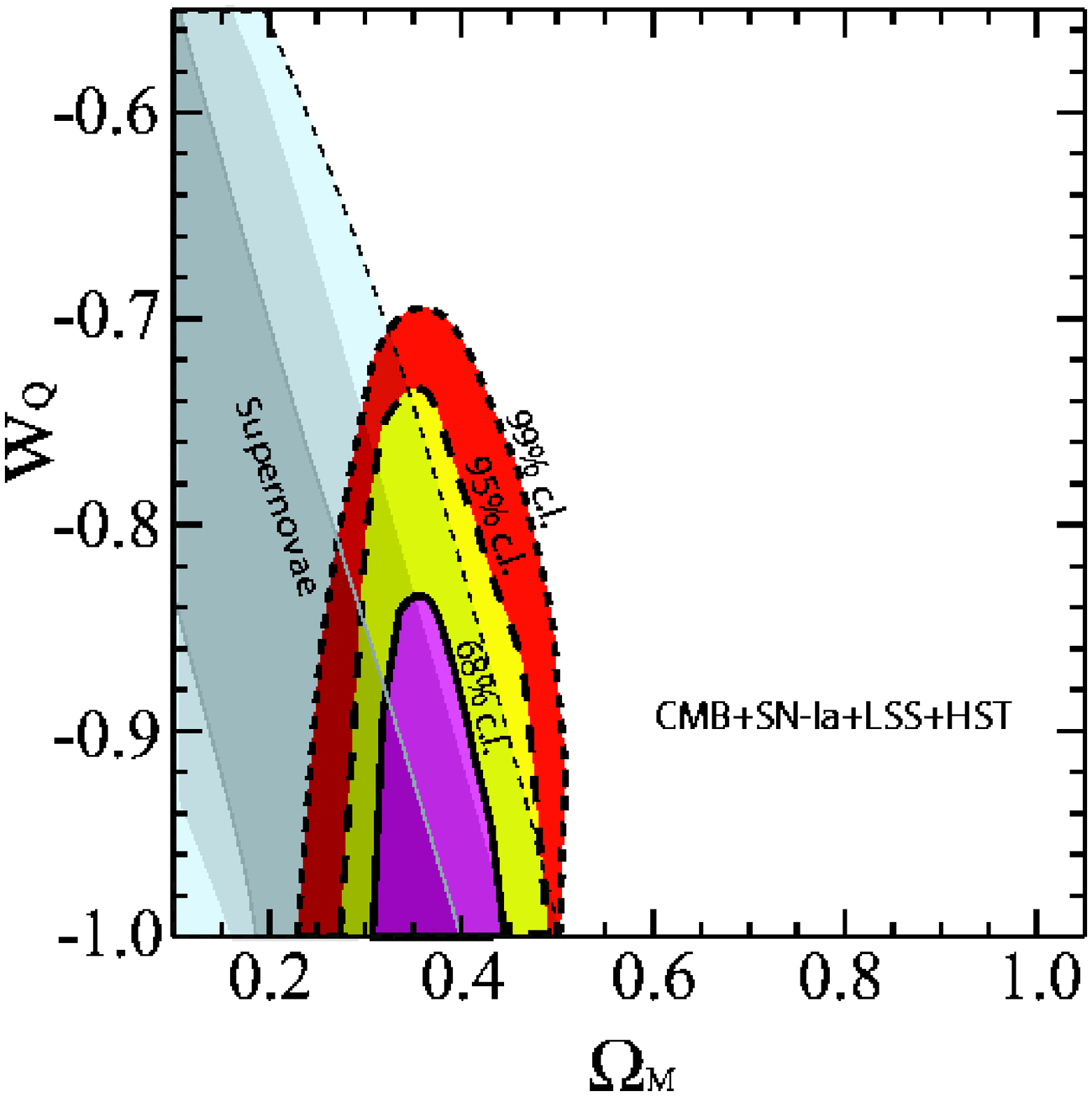}
\end{center}
\caption{The likelihood contours in the ($\Omega_M$, $w_Q$) plane,
with the remaining parameters taking their best-fitting values for the
joint CMB+SN-Ia+LSS analysis described in the text.
The contours correspond to 0.32, 0.05 and 0.01 of the peak value of the
likelihood, which are the 68\%, 95\% and 99\% confidence levels respectively.
Picture taken from \cite{rachel}.}
\label{figo1}
\end{figure}
\medskip                                                                      

In Figure 7 we plot likelihood contours in the ($\Omega_M$, $w_Q$) plane 
for the joint analyses of CMB+SN-Ia+HST+LSS data together with 
the contours from the SN-Ia dataset only. 
Proceeding as in \cite{melchiorri}, we attribute a likelihood to a point
in the ($\Omega_M$, $w_Q$) plane by finding the remaining parameters that
maximize it. We then define our $68\%$, $95\%$
 and  $99\%$ contours to be where the likelihood falls to $0.32$, $0.05$ and
$0.01$ of its peak value, as would be the case for a
two dimensional multivariate Gaussian. 
As we can see, the combination of the datasets breaks the
luminosity distance degeneracy and suggests the presence of dark
energy with high significance. 
Furthermore, the new CMB results provided by Boomerang and DASI improve 
the constraints 
from previous and similar analysis (see e.g., \cite{PTW},\cite{bondq}), with
$w_Q<-0.85$ at $68 \%$ c.l..
Our final result is then perfectly in agreement with the $w_Q=-1$
cosmological constant case and gives no support to a
quintessential field scenario with $w_Q > -1$.

\subsection{CMB, Big Bang Nucleosynthesis and Neutrinos}

As we saw in the previous section, the SBBN $95 \%$ CL region,
corresponding to $\Omega_b h^2= 0.020 {\pm} 0.002$ ($95 \%$ c.l.), 
has a large overlap with the analogous CMBR contour. 
This fact, if it will be confirmed by future experiments on CMB
anisotropies, can be seen as one of the greatest
success, up to now, of the standard hot big bang model.

SBBN is well known to provide strong bounds on the number 
of relativistic species $N_\nu$. On the other hand,
Degenerate BBN (DBBN), first analyzed in Ref. \cite{d1,d2,d3,Kang}, gives
very weak constraint on the effective number of massless neutrinos, since
an increase in $N_\nu$ can be compensated by a change in both the chemical
potential of the electron neutrino, $\mu_{\nu_e}= \xi_e T$,
and $\Omega_bh^2$. 
Practically, SBBN relies on the theoretical assumption that 
background neutrinos have negligible chemical potential, just like their 
charged lepton partners. Even
though this hypothesis is perfectly justified by Occam razor, models have
been proposed in the literature
\cite{dibari,AF,DK,DolgovRep,Casas,MMR,McDonald,Foot}, where large neutrino
chemical potentials can be generated. It is therefore an interesting issue
for cosmology, as well as for our understanding of fundamental
interactions, to try to constrain the neutrino--antineutrino asymmetry
with the cosmological observables. It is well known that degenerate BBN gives
severe constraints on the electron neutrino chemical potential, $-0.06\leq
\xi_e\leq 1.1$, and weaker bounds on the chemical potentials of both 
the $\mu$ and $\tau$
neutrino, $|\xi_{\mu,\tau}|
\leq 5.6 \div 6.9$ \cite{Kang}, since electron neutrinos are directly
involved in neutron to proton conversion processes which eventually fix the
total amount of $^4He$ produced in nucleosynthesis, while $\xi_{\mu,\tau}$
only enters via their contribution to the expansion rate of the universe.

The CMB power spectrum is greatly affected by changes in $N_{\nu}$, 
the amount of relativistic particles.
First of all, changing $N_{\nu}$ changes the epoch of equality.
Secondly, the shift parameter $\cal R$ is changed as
(\cite{bowen}):

\be
\label{eq:def_r}
{\cal R} = 2 \left( 1 - \frac{1}{\sqrt{1 + z_{dec}}} \right) 
\frac{\sqrt{| \Omega_k| }}{ \Omega_m} \frac{1}{\chi(y)}
    \left[ \sqrt{\Omega_{rel} +
    \frac{\Omega_m}{1 + z_{dec}}} - \sqrt{\Omega_{rel}} \right]
\ee

where now

\be
y = \sqrt{|\Omega_k|}\int_0^{z_{dec}}dz{[\Omega_{rel} (1+z)^4 +
\Omega_m(1+z)^3+\Omega_k(1+z)^2+\Omega_{\Lambda}]^{-1/2}}. \nonumber
\ee                                                               
  
Finally, in the acoustic peaks region,
the different radiation content at decoupling 
by variation of the ratio $\Omega_{\gamma}/\Omega_{rel}$ 
induces a larger early ISW effect, which boosts
the height of the first peak with respect to the other
acoustic peaks. 

Combining the DBBN scenario with the bound on baryonic and radiation densities
allowed by CMBR data, it is possible to obtain stronger constraints
on all the parameter of the model. 
Such an analysis was previously performed in
(\cite{th7}, \cite{peloso}, \cite{hannestad}, \cite{orito}) 
using the first data release of BOOMERanG and MAXIMA 
(\cite{debe00}, \cite{hanany}). 
We recall that the neutrino chemical potentials contribute
to the total neutrino effective degrees of freedom $N_\nu$ as

\begin{equation}
N_{{\nu}} = 3 + \Sigma_{\alpha} \left[ \frac{30}{7}
\left( \frac{\xi_\alpha}{\pi} \right)^2 +
\frac{15}{7} \left( \frac{\xi_\alpha}{\pi} \right)^4 \right] \, .
\end{equation}

Notice that in order to get a bound on $\xi_\alpha$ we have here assumed   
that all relativistic degrees of freedom, other than photons, are given by
three (possibly) degenerate active neutrinos.

\begin{figure}
\begin{center}
%\epsfxsize=6.2cm
%\epsfysize=6.2cm
%\epsffile{cont.ps}
\includegraphics[scale=0.4]{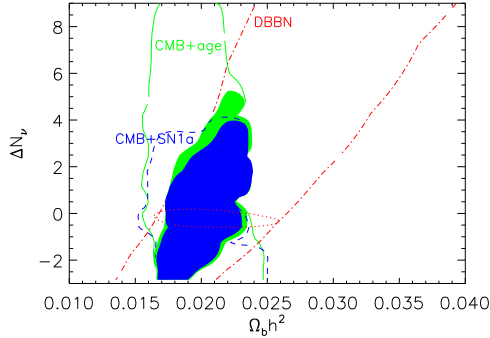}
\end{center}
\caption{The $95\%$ CL contours for degenerate BBN (dot-dashed (green) line),
new CMB results with just the age prior, $t>11$gyr (full (red) line),
and with just the SN1a prior (dashed (blue) line).
The combined analysis corresponds to the filled regions.
Marginalization leads to the bound
$\Omega_b h^2 =  0.020{\pm} 0.0035$ and
$N_\nu < 7$, both at $95 \%$, for DBBN+CMB+SN.
The dotted (green) line is the $95 \%$ CL allowed by SBBN.
Picture taken from \cite{hmmmp}.}
\label{figDbbn}
\end{figure}

Figure~8 summarizes our main results with the new CMB data, reported in
\cite{hmmmp} for the DBBN scenario. We plot the 
$95\%$ CL contours allowed by DBBN (dot-dashed (green) line),
together with the analogous $95 \%$ CL 
region coming from the CMB data analysis,
with only weak age prior, $t_0 > 11 $gyr (full (red) line).

Finally, the solid contour (light, red) is
the $95 \%$ CL region of the joint product distribution ${\cal L} \equiv
{\cal L}_{DBBN}$${\cdot}{\cal L}_{CMB}$.
The main new feature, with respect to the results
of Ref. \cite{th7} is that the resolution of the third peak shifts the CMB
likelihood contour towards smaller values for $\Omega_b h^2$, so when
combined with DBBN results, it singles out smaller values for $N_\nu$. In
fact from our analysis we get the bound $N_\nu \leq 8$, at $95 \%$ CL,
which translates into the new bounds $-0.01\leq \xi_e \leq 0.25$, and
$|\xi_{\mu,\tau}|\leq 2.9$, sensibly more stringent than what can be
found from DBBN alone.

A similar analysis can also be performed combining CMBR and DBBN data with
the Supernova Ia data \cite{super1}, which strongly reduces the degeneracy
between $\Omega_m$ and $\Omega_\Lambda$. At $95 \%$ C.L. we find
$\Delta N_\nu < 7 $, corresponding to  $-0.01\leq \xi_e \leq 0.22$ and
$|\xi_{\mu,\tau}|\leq 2.6$.  

Compatible results have been obtained in 
similar analyses (\cite{kneller},\cite{hannestad2}).

Some caution is naturally necessary when comparing the effective number of
neutrino degrees of freedom from BBN and CMB, since they may be related to
different physics. In fact the energy density in relativistic species may
change from the time of BBN ($T \sim MeV$) to the time of last rescattering 
($T\sim eV$). It is therefore interesting to further investigate
the bounds on this parameter from CMB alone and see how the inclusion
of a free $N_\nu$ can affect the results on parameter extraction.

\begin{figure}
\begin{center}
%\epsfxsize=7.2cm
%\epsfysize=6.2cm
%\epsffile{fig1.eps}
\includegraphics[scale=0.25]{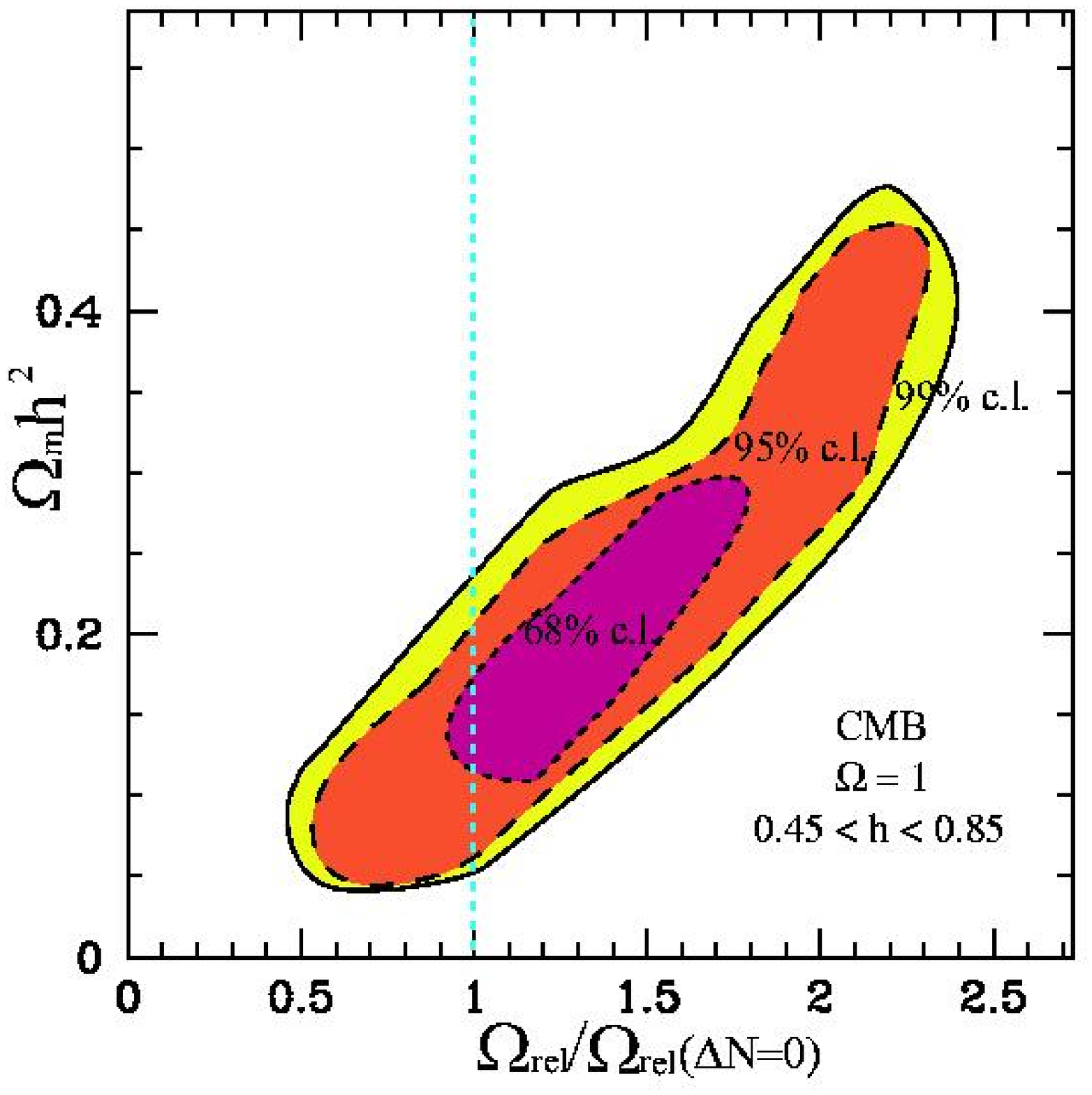}
\includegraphics[scale=0.25]{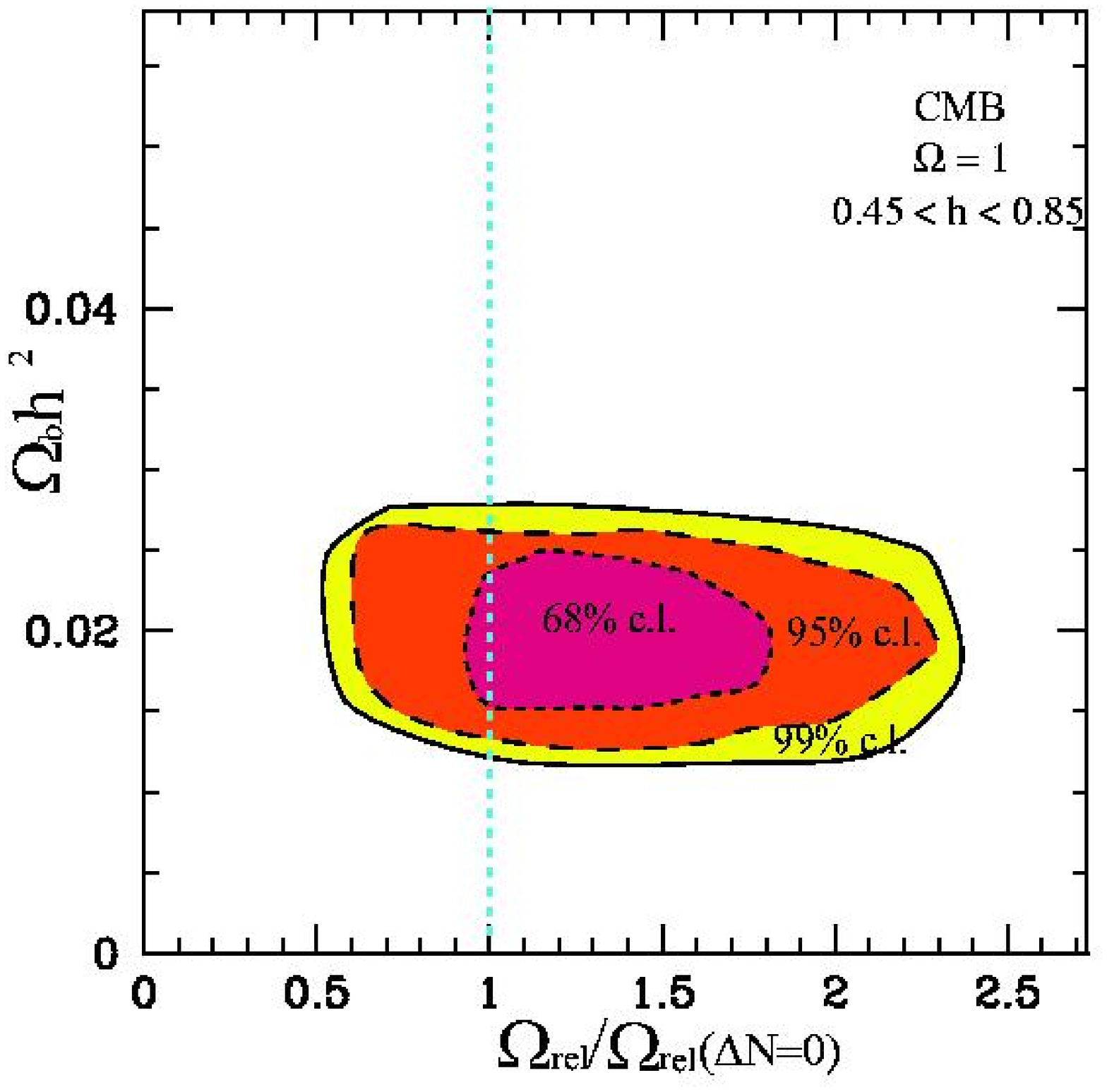}                              
\includegraphics[scale=0.25]{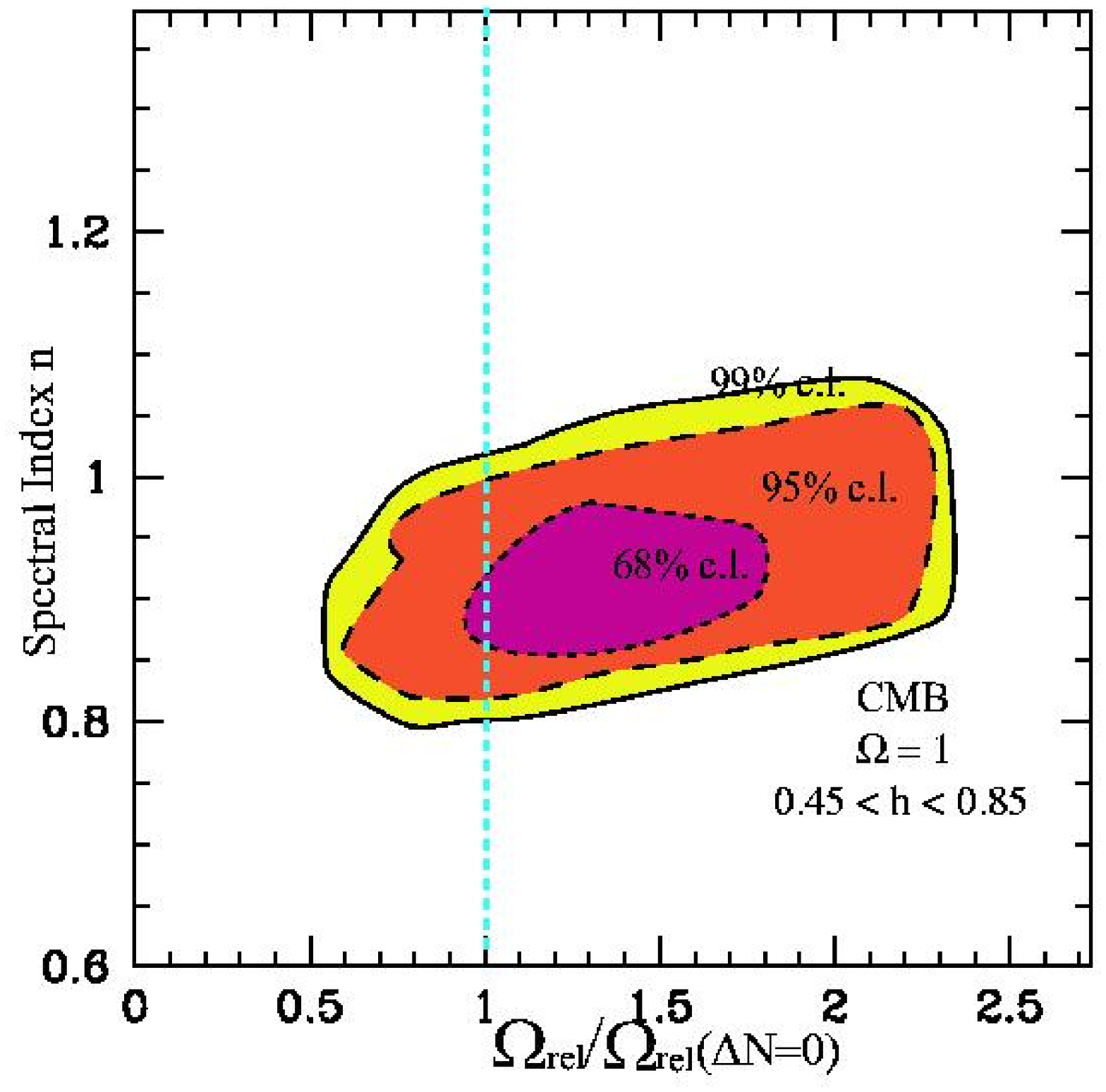}                                   
\end{center}
\caption{Effects of a background of relativistic particles on the
parameters estimated from CMB observations. 
Picture taken from \cite{bowen}.}
\label{fig9}
\end{figure} 

We have analyzed this in \cite{bowen}. 
In Figure 9 we plot the likelihood contours for $\omega_{rel}$ vs
$\omega_m, \omega_b$ and $n_s$ (left to right).  As we can see,
$\omega_{rel}$ is weakly constrained to be in the range $1 \le
\omega_{rel}/\omega_{rel}(\Delta N_{\nu}=0) \le 1.9$ at $1-\sigma$ in all the
plots. The degeneracy between $\omega_{rel}$ and $\omega_{m}$ is
evident in the left panel of Figure 9.  Increasing $\omega_{rel}$
shifts the epoch of equality and this can be compensated only by a
corresponding increase in $\omega_m$. It is interesting to note that
even if we are restricting our analysis to flat models, the degeneracy
is still there and that the bounds on $\omega_m$ are strongly
affected.  We find $\omega_m=0.2\pm0.1$, to be compared with
$\omega_m=0.13\pm0.04$ when $\Delta N_{\nu}$ is kept to zero.  
It is important to realize that these bounds on $\omega_{rel}$ appear 
because of our prior on $h$ and because we consider flat models. 
When one allows $h$ as a free parameter and any value for $\Omega_m$, 
then the degeneracy is almost complete. 
In principle, a critical univers with $\Omega_M=1$ could be
put back in agreement with CMB and LSS observations by
this mechanism (\cite{lesg}).
In the center and right panel of Fig.9 we plot 
the likelihood contours for $\omega_b$ and $n_s$.  
As we can see, these parameters are not
strongly affected by the inclusion of $\omega_{rel}$ and the
most relevant degeneracy is with the amount of
non relativistic matter $\omega_m$.
An accurate determination of $\omega_{cdm}=\omega_m-\omega_b$ can shed new
light on the nature of dark matter. The thermally averaged
cross-sections times velocity of the dark matter candidate is related
to $\omega_{cdm}$, and this relation is currently used to analyze the
implications for the mass spectra in versions of the Supersymmetric
Standard Model (see e.g. \cite{Barger}, \cite{djouadi}, \cite{ellis}).

\subsection{Varying $\alpha$}

\begin{figure}[ht]
\begin{center}
\includegraphics[scale=0.42]{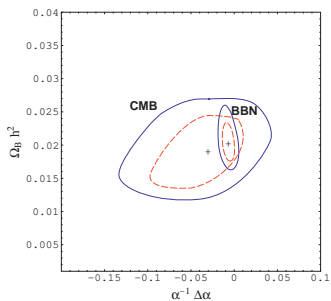}
\end{center}
\caption{BBN and CMB likelihood contours for variations in the fine
structure constant $\alpha$. Picture taken from \cite{avelino}}
\label{figo}
\end{figure}
\medskip

There are quite a large number of experimental constraints on the value of
fine structure constant $\alpha$. 
These measurements cover a wide range of timescales (see
\cite{alpharev} for a review of this subject), starting from present-day
laboratories ($z \sim 0$), geophysical tests ($z << 1$), and quasars
($z\sim 1 \div 3$), through the CMB ($z\sim 10^3$) and BBN ($z\sim10^{10}$)
bounds.

The recent analysis of \cite{Webb} of fine splitting of quasar doublet 
absorption lines gives a $4\sigma$ evidence for a time variation 
of $\alpha$, $\Delta \alpha/\alpha=(-0.72 {\pm} 0.18) 10^{-5}$, 
for the redshift range $z \sim 0.5 - 3.5$. 
This positive result was obtained using a many-multiplet
method, which, it is claimed, achieves an order of magnitude greater precision
than the alkali doublet method. Some of the initial ambiguities of the method
have been tackled by the authors with an improved technique, in which a
range of ions is considered, with varying dependence on $\alpha$, which
helps reduce possible problems such as varying isotope ratios, calibration
errors and possible Doppler shifts between different populations of ions
\cite{Quas1,Quas2,Quas3,Quas4}.

The present analysis of the $\alpha$-dependence of two relevant cosmological
observables like the anisotropy of CMB and the light element primordial
abundances does not support evidence for variations of the fine-structure
constant at more than the one-sigma level at either epoch.
The $68 \%$ and $95 \%$ C.L. regions in the plane 
$\Omega_bh^2$-- $\Delta\alpha / \alpha$ are shown in Fig. 10 from CMB
and BBN (see \cite{avelino} and references therein).

\subsection{Isocurvature modes.}

Another key assumption is that the primordial fluctuations were adiabatic.
Adiabaticity is not a necessary consequence of inflation though and 
many inflationary models have been constructed where
isocurvature perturbations would have generically been concomitantly
produced (see e.g. \cite{langlois}, \cite{gordon}, \cite{bartolo}).

In a phenomenological approach one should consider the most
general primordial perturbation, introduced by \cite{kavi}, and
described by a $5X5$ symmetric matrix-valued generalization of
the power spectrum.
As showed by \cite{kavi}, the inclusion of isocurvature perturbations with 
auto and cross-correlations modes has dramatic
effects on standard parameter estimation with uncertainties 
becoming of order one.

Even assuming priors such as flatness, the inclusion of isocurvature
modes significantly enlarges our constraints on the baryon density
\cite{trotta} and the scalar spectral index \cite{amendola}.
Pure isocurvature perturbations are highly excluded by present CMB
data (\cite{enq}).

As we saw in the first section, it is also possible to have {\it active} and 
{\it decoherent} perturbations such as those produced by an inhomogeneously
distributed form of matter like topological defects.
Models based on global defects like cosmic strings and textures are
excluded at high significance by the present data (see e.g. \cite{DKMrep}).
However a mixture of adiabatic$+$defects is still compatible with the
observations (\cite{bouchet}, \cite{DKMrep}).
In principle, toy models based on {\it active} perturbations can
be constructed \cite{turok} that can mimic inflation and retain
a good agreement with observations \cite{dkm2}.

\section{Conclusions}

The recent CMB data represent a beautiful success for the 
standard cosmological model. The acoustic oscillations in
the CMB angular power spectrum, a major
prediction of the model, have now been detected at 
$\sim 5 \sigma$ C.L. for the first peak and $\sim 2 \sigma$ C.L.
for the second and third peak.
Furthermore, when constraints on cosmological parameters are 
derived under the assumption of adiabatic primordial perturbations 
the following results are obtained:

\begin{itemize}

\item The curvature of the universe is zero, i.e. the universe
is flat, in agreement with the predictions of the theory.

\item The power spectrum of the primordial perturbations is
nearly scale-invariant, again a prediction of the model.

\item The amount of density in baryons is in agreement with
independent observations of primordial abundances and
 standard big-bang nucleosynthesis.

\item The optical depth is constrained to be $\tau_c <0.3$,
and the universe recombined, in agreement with the 
overall scenario.

\item Some form of non baryonic dark matter must be present,
as requested by a large set of independent observations.

\item The age of the universe is consistent with at least 
$3$ independent constraints.

\item When information from complementary datasets, like
constraint on $h$ or from large scale structure, are included
in the analysis, the CMB data suggest a presence for a cosmological
constant in agreement with the SN-Ia result.

\end{itemize}

All these results strongly suggest that the inflationary 
scenario of structure formation is coherent in its simplest form.
In few words, the hoped-for miracle.

As we saw in the previous section, modifications to this model, 
like adding isocurvature modes or topological defects, are in
agreement with the observations, but are not required by the data and
are reasonably constrained when complementary datasets are included
in the analysis.

Since the model is in agreement with the data and all the most relevant
parameters are starting to be constrained within a few percent accuracy,
the CMB is becoming a wonderful laboratory for investigating the 
possibilities of new physics. With the promise of 
large data sets from Map, Planck and SNAP satellites, 
opportunities may be open, for example, to constrain dark energy models, 
variations in fundamental constants and neutrino physics.

%%
%%   use this format to include an .eps figure into your paper
%%
%\begin{figure}[htb]
%    \centering
%    \includegraphics[height=1.5in]{helicity.eps}
%    \caption{Figure caption.}
%    \label{fig:cosmo}
%\end{figure}
%%%%%%%%%%%%%%%%%%%%%%%%%%%%%%%%%%%%%%%%%%%%%%%%%%%%%%%%%%%%%%%%%%%%%%%%

\Acknowledgements

First of all, I wish to thank all the organizers of this
exciting conference: Matts Roos, Kari Enqvist, 
Hannu Kurki-Suonio, Anna Kalliomaki
Antti Sorri and Jussi Valiviita in particular.
I am grateful to Becky Bowen and Steen Harle Hansen for 
comments and suggestions.
Many thanks also to Rachel Bean, Celine Boehm, Sarah Bridle, 
Rob Crittenden, Ruth Durrer, Pedro Ferreira, Ignacio Ferreras, Will Kinney, 
Gianpiero Mangano, Carlos Martins, Gennaro Miele, Marco Peloso, 
Ofelia Pisanti, Antonio Riotto, Graca Rocha, Joe Silk, 
and Roberto Trotta for comments, discussions and help.

\end{document}